\documentclass[aps,prc,showpacs,nofootinbib]{revtex4}

\usepackage{epsfig}
\usepackage{graphicx}

\begin{document}
\title{Exclusive photoproduction of charmed $D^*$ vector mesons, 
$\gamma+N\to Y_c+ \overline{D}^*$, $Y_c=\Lambda_c^+$ or  $\Sigma_c$}
\author{Egle Tomasi-Gustafsson}
\email{etomasi@cea.fr}
\affiliation{\it DAPNIA/SPhN, CEA/Saclay, 91191 Gif-sur-Yvette Cedex,
France}
\author{Michail P. Rekalo} 
\altaffiliation{Permanent address:
\it NSC Kharkov Institute of Physics and Technology, 61108 Kharkov, Ukraine}
\affiliation{Middle East Technical University,
Physics Department, Ankara 06531, Turkey}

\date{\today}

\begin{abstract}
The spin structure of the matrix element for the reactions $\gamma+ N\to Y_c +{D}^*$, where $Y_c=\Lambda_c^+(2285)$, $\Sigma_c(2455)$ are charmed baryons with spin 1/2, and $D^*(2010)$ is the vector charmed meson, can be parametrized, in collinear regime, in terms of three independent scalar amplitudes, which are functions of the photon energy $E_{\gamma}$, only.  In framework of an effective Lagrangian approach generalized to charm photoproduction, we calculate the energy dependence of the differential cross section, the density matrix element of $\overline{D}^*$, $\rho_{11}$, the asymmetry $A_z$ in the collision of circularly polarized photons with polarized nucleons, and the polarization of the produced $Y_c$-hyperon, $P_z$, in the collision of circularly polarized photons with unpolarized target. All these polarization observables either vanish or are large, in absolute value, with a smooth $E_{\gamma}$-dependence, and differ for $\Lambda_c^+$ and $\Sigma_c$ --production.
\end{abstract}

\maketitle

\section{Introduction}
In a previous paper \cite{Etg03} we analyzed the exclusive processes of charm pseudoscalar meson ${D}^0$ photoproduction,
$\gamma+N\to Y_c+ \overline{D}$, $Y_c=\Lambda_c^+$, or  $\Sigma_c$. We consider here the exclusive processes of charmed vector meson ${D}^*$ photoproduction,  
$\gamma+N\to Y_c+ \overline{D}^*$, on proton and neutron targets. No detailed theoretical analysis exists for such processes, and up to now, experimentally, only an indirect evaluation of their cross section has been done. In \cite{Ma99} an attempt has been done to estimate  the contribution of the processes $\gamma+N\to Y_c+ \overline{D}(\overline{D}^*)$ to the asymmetry of the collisions of a circularly polarized photon beam with a polarized target in $\vec\gamma+\vec N\to$ open charm $+X$. It is well known, that such asymmetry is sensitive to the $\Delta G $ gluon contribution to the nucleon spin \cite{PGF}. In \cite{Re01} the D-exchange contribution has been calculated for 
$\gamma + N\to Y_c+ \overline{D}^*$, in the near threshold region, considering the possible baryon exchange as a background process.

Let us list some arguments to justify the interest in exclusive charmed vector meson photoproduction:
\begin{itemize}
\item their contribution to the total cross section is important, especially in the near-threshold region;
\item the understanding of the reaction mechanisms require the deconvolution of these channels;
\item these reactions naturally explain the particle-antiparticle asymmetry in charmed particles photoproduction, at high energies;
\item the polarization phenomena induced by the production of vector meson are new, interesting and measurable (the $D^*$-meson is a self analyzing particle, through the angular dependence of the decay products in $D^*\to D+\pi$, but this decay does not allow to access the vector $D^*$-meson polarization);
\item direct experimental data, concerning these reactions, are expected soon, due to the higher luminosity of ongoing experiments, as COMPASS \cite{COMPASS}.
\end{itemize}

There are essential differences beween the processes of exclusive  photoproduction of light vector mesons ($\rho$, $\omega$, $\phi$ or $K^*$) and heavy $D^*$-production.  In the last case, the diffractive mechanism is absent and the large mass of the $c$-quark seems to justify the applicability of perturbative QCD. However there is no direct relation between the photon-gluon fusion subprocess, $\gamma+G\to c+\overline{c}$  and the exclusive $D^*$-photoproduction. Therefore, QCD-inspired models, like the effective Lagrangian approach (ELA), in terms of hadronic degrees of freedom, as charmed baryons and mesons, seem more appropriate. In such approach, it is possible to predict the differential cross section and all polarization observables in terms of a finite number of coupling constants, of strong and electromagnetic nature - in any kinematical condition. The same coupling constants enter also in the calculation of other processes of associative charm production in $\pi N$, $\gamma N$ and $NN$-collisons.

There are experimental data about inclusive $D^*$ photoproduction, $\gamma+p\to D^*+X$ or $e^-+N\to e^-+D^*+X$ up to HERA energies 
\cite{HERA}. The smallest energy, where $D^*$-photoproduction has been observed is $E_{\gamma}=20$ GeV, at SLAC \cite{Ab86}. Typically, $D^*$-production is the main channel for $\gamma+ N\to$ charm $+X$ \cite{HERA,Ab86,Ad87,Al92,Anjos89,Sl83,Li03}, so that the ratio between the vector ($V$) and the total $D^*$ contribution to the total cross section is in agreement with the spin quark rule $V/(P+V)\simeq 0.75$, where $P$ is the pseudoscalar contribution. The relative role of the charged and neutral $D$-meson photoproduction depends, on one side, on the relative cross sections for  the $D^{*\pm}$ and $D^{*0}$ photoproduction, and, from another side, on the branching ratio for the decays $D^*\to D+\pi $ (for different charge combinations). 

We consider here the exclusive processes 
$\gamma+N\to Y_c+ \overline{D}^*$, in case of collinear kinematics, where helicity conservation and the P-invariance of electromagnetic interaction allow three amplitudes, only, which are functions of a single kinematical variable, the photon energy $E_{\gamma}$. Note, in this respect, that generally the processes $\gamma+N\to Y_c+ \overline{D}^*$ are characterized by a set of twelve independent amplitudes, which are complex functions of two kinematical variables. Therefore the theoretical analysis is largely simplified in collinear kinematics.

This paper is organized as follows. In Section II the spin structure of the collinear matrix element is parametrized in terms of three amplitudes, which allows to develop a general and model independent formalism for the analysis of polarization phenomena. The expressions for the different possible pole contributions, in framework of ELA approach, are derived in Section III. The strong and electromagnetic coupling constants as well as the form factors are discussed in Section IV. Numerical predictions for the differential cross section, the single and double-spin polarization observables, as functions of $E_{\gamma}$, are given in  Section V. Concluding remarks are given in Section VI. The Appendix contains the explicit expressions of the collinear amplitudes.

\section{Collinear amplitudes and polarization phenomena}

We consider here the processes $\gamma+N\to Y_c+ \overline{D}^*$, $Y_c=\Lambda_c^+$, or  $\Sigma_c$, $\overline{D}^*$=$\overline{D}^{*0}$ or $\overline{D}^{*-}$, in collinear regime, i.e. for $\theta=0$ or $\pi$, where $\theta$ is the $\overline{D}^{*-}$-production angle in the reaction CMS. The reasons of this choice are the following:
\begin{itemize}
\item the differential cross section, $d\sigma/dt$, as a rule, is maximal in the forward direction ($t$ is the momentum transfer squared);
\item the total helicity of the interacting particles is conserved, for any reaction mechanisms, which implies that the spin structure of the matrix element
and the polarization phenomena are highly simplified.
\end{itemize}

At high photon energies, collinear kinematics is very near to the condition of forward detection of the running experiments, such as COMPASS \cite{COMPASS}.

Taking into account the P-invariance of the electromagnetic interaction of charmed particles and the helicity conservation, one can write the following general parametrization of the matrix element for any process $\gamma +N\to Y_c+\overline{D}^*$, in collinear kinematics (which holds for any reaction mechanism) :
\begin{equation}
{\cal M}(\gamma N\to Y_c\overline{D}^*)=\chi_2^{\dagger}\left 
[\vec \epsilon\cdot\vec {U^*} f_1 +i 
\vec\sigma\cdot\hat{\vec k}\vec \epsilon \times \hat{\vec k}\cdot\vec {U^*} f_2
+i\vec\sigma\cdot\vec \epsilon\times \hat{\vec k } \vec {U^*} \cdot\hat{\vec k} f_3\right ]\chi_1,
\label{eq:mat}
\end{equation}
with the following notations:
\begin{itemize}
\item  $\chi_1$ and  $\chi_2$ are the two-component spinors of the initial nucleon and the produced $Y_c$-hyperon,
\item $\vec \epsilon  $ and $\vec {U}$ are the three vector of the photon and of the $\overline{D}^*$-meson polarizations, with the condition 
$\epsilon\cdot \hat{\vec k }=0$,
\item 
$\hat{\vec k}$ is the unit vector along the three-momentum of $\gamma$; 
\item $f_i$, $i$=1-3, are the collinear amplitudes, which are generally complex functions of a single variable, $E_{\gamma}$.
\end{itemize}
Using this parametrization, one can find for the differential cross section:
\begin{equation}
\displaystyle\frac{d\sigma}{dt}={\cal N}\left [ |f_1|^2+|f_2|^2+\displaystyle\frac{E_v^2}{m_v^2}|f_3|^2 \right ],
\label{eq:cs1}
\end{equation}
where ${\cal N}$ is a normalization factor, $E_v$ $(m_v)$ is the energy (mass) of the produced vector meson:
\begin{equation}
E_v=\displaystyle\frac{s+m_v^2-M^2}{2W},~s=W^2=m^2+2mE_{\gamma},
\label{eq:cs}
\end{equation}
where $m$ $(M)$ is the nucleon ($Y_c$-hyperon) mass, $E_{\gamma}$ is the photon energy in the laboratory (Lab) system.

The $D^{*}$-mesons, produced in collinear kinematics, are generally polarized, (with tensor polarization) even in collisions of unpolarized particles:
\begin{equation}
{\cal D}\rho_{11}=\displaystyle\frac{|f_1|^2+|f_2|^2}{2},~
{\cal D}=|f_1|^2+|f_2|^2+\displaystyle\frac{E_v^2}{m_v^2}|f_3|^2
\label{eq:rho}
\end{equation}
with the normalization condition: $2\rho_{11}+\rho_{00}=1$. The non-diagonal elements for $\rho_{ab}$ are equal to zero, in collinear regime.

All the other single-spin polarization observables, such as the $\Sigma_B$-asymmetry (with linearly polarized photons interacting with unpolarized target), the analyzing power $A$ (induced by polarized nucleon target) and the final $Y_c$-polarization (in the collision of unpolarized particles), vanish for the considered reactions, for any photon energy and for any reaction mechanism, due to the axial symmetry of collinear kinematics. However an interesting set of double-spin polarization observables can be measured, for the reactions  $\gamma+N\to Y_c+ \overline{D}^*$:
\begin{itemize}
\item {\bf the asymmetry $A_z$} in the collision of circularly polarized photon beam, with a polarized nucleon target, in the $\hat{\vec k}$-direction, which we choose as the $z$-direction, 
\begin{equation}
A_z{\cal D}=-2Re ~f_1f_2^*-|f_3|^2\displaystyle\frac{E_v^2}{m_v^2}.
\label{eq:az}
\end{equation}
Due to the P-invariance of the electromagnetic interaction, the asymmetry for a linearly polarized photon beam vanish, even in collisions with polarized target. In the same way,  the linear photon polarization can not induce any polarization of the emitted $Y_c$-hyperon (for simplicity, we will assume here 100\% photon polarization, with helicity $\lambda_{\gamma}=+1$). 
\item In the collision of circularly polarized photons with unpolarized target, the $Y_c$-hyperon can be longitudinally polarized, and {\bf the polarization $P_z$}  is:
\begin{equation}
P_z{\cal D}=-2Re ~f_1f_2^*+|f_3|^2\displaystyle\frac{E_v^2}{m_v^2}.
\label{eq:pz}
\end{equation}
\item 
The nonzero components of {\bf the depolarization tensor, $D_{ab}$}, describing the dependence of the  $b$-component of the $Y_c$-polarization on the $a$-component of the target polarization can be written as:
\begin{equation}
D_{zz} {\cal D}=|f_1|^2+|f_2|^2 -|f_3|^2\displaystyle\frac{E_v^2}{m_v^2},
~D_{xx} {\cal D}=D_{yy}{\cal D}=|f_1|^2-|f_2|^2.
\label{eq:dab}
\end{equation}
\end{itemize}
One can see that these observables are not independent, and the following relations hold, at any photon energy and for any reaction mechanism:
\begin{equation}
D_{zz}=-1+4\rho_{11},~A_z-P_z=-2+4\rho_{11}.
\label{eq:rel}
\end{equation}
These formulas show which experiments are necessary to determine the moduli of the collinear amplitudes, $|f_i|$:
\begin{eqnarray}
|f_1|^2&=&(\rho_{11}+ \displaystyle\frac{1}{2}D_{xx}){\cal D},
\nonumber\\
|f_2|^2&=&(\rho_{11}-\displaystyle\frac{1}{2}D_{xx}){\cal D},\label{eq:cam}\\
|f_3|^2\displaystyle\frac{E_v^2}{m_v^2}&=  & (1-2\rho_{11}){\cal D}.\nonumber 
\end{eqnarray}
Therefore, the measurements of $\rho_{11}$ and $D_{xx}$, together with the differential cross section $d\sigma/dt$ can be considered as the first step of the complete experiment for any collinear reaction of vector meson photoproduction on nucleon, such as, for example:
\begin{eqnarray}
&\gamma+N\to N+V,&~V=\rho,\omega,\phi,\nonumber \\
&\gamma+N\to Y+K^*,&~Y=\Lambda \mbox{~or~}\Sigma- \mbox{~hyperon}\label{eq:reac}\\
&\gamma+N\to Y_c+\overline {D}^*.&\nonumber
\end{eqnarray}

Further experiments are necessary to determine the relative phases of the collinear amplitudes $f_i$. For example, the relation:
\begin{equation}
(P_z+A_z){\cal D}=-4 Re f_1f_2^*
\label{eq:rel1}
\end{equation}
allows to determine the relative phase, $\delta_1-\delta_2$, of the amplitudes $f_1$ and $f_2$, more exactly, $\cos(\delta_1-\delta_2)$. T-odd polarization observables are more sensitive to the small relative phase, being determined by $\sin(\delta_1-\delta_2)$. The simplest of these observables, in collinear regime, is the tensor $D^*$-polarization, induced by polarized target. The corresponding density matrix can be parametrized as follows:
\begin{equation}
\rho_{ab}(P)=i\rho_1\epsilon_{abc}P_c+i\rho_2\epsilon_{abc}
\hat k_c \hat{\vec k}\cdot\vec P+\rho_3[\hat k_a (\hat{\vec k}\times\vec P)_b+
\hat k_b (\hat{\vec k}\times\vec P)_a],
\label{eq:rh}
\end{equation}
where $\rho_i$, $i=1-3$, are real coefficients, quadratic functions of the collinear amplitudes and $\vec P$ is the pseudovector of the target polarization:
\begin{eqnarray}
&\rho_1{\cal D}=  & Re ~f_1f_3^*, \nonumber\\
&\rho_2{\cal D}= &Re ~(-f_1f_2^*-f_1f_3^*+f_2f_3^*),
\label{eq:camr}\\
&\rho_3 {\cal D}=&Im~(f_1-f_2)f_3^*.\nonumber
\end{eqnarray}

Note, in this connection, that the antisymmetrical part of $\rho_{ab}(P)$, which is characterized by the coefficients $\rho_1$ and $\rho_2$, describes  $D^*$-production with vector polarization. But this polarization can not be measured through the main decays of $D^*$: $D^*\to D +\pi$, $D^*\to D+\gamma$, and $D^*\to D+e^++e^-$ \cite{Al94}. The coefficient $\rho_3$ characterizes the dependence of the $D^*$-tensor polarization on the target polarization, generating the following angular distribution for the decay products in $D^*\to D +\pi$:
\begin{equation}
W(\theta,\phi)\simeq\sin 2\theta\sin\phi,
\label{eq:ww}
\end{equation}
where $\theta$ and $\phi$ are the polar and azimuthal angles of the $\pi$-meson (in  $D^*$-rest frame) relative to the polarization plane, which is defined by $\hat{\vec k}$ and $\vec P$. 

Linear photon polarization can also induce polarized $D^*$-mesons, with the following non zero elements of density matrix:
\begin{eqnarray}
\rho_{11}^{(1)}{\cal D}&=  & \displaystyle\frac{1}{2}(|f_1|^2+|f_2|^2), \nonumber\\
\rho_{1-1}^{(1)}{\cal D}&= &\displaystyle\frac{1}{2}(|f_1|^2-|f_2|^2),
\label{eq:camr1}
\end{eqnarray}

This represents the complete analysis of all possible double-spin polarization observables, for the reactions $\gamma+N\to Y_c+\overline{D}^*$, in collinear kinematics.

\section{The matrix elements}
In a previous work \cite{Etg03}, we considered the processes of photoproduction of pseudoscalar $\overline{D}$-mesons, in $\gamma+N\to Y+\overline{D}$. In this section we analyze the processes $\gamma+N\to Y_c+\overline{D}^*$, following a similar scheme. These two classes of reactions have common properties with the reactions of associative strange particles production, $\gamma+N\to Y+K$ and $\gamma+N\to Y+K^*$, where $Y=\Lambda$ and $\Sigma$, strange hyperons, even if the masses of the produced particles are very different. All these processes have a non diffractive nature - due to the quantum numbers in $t-$channel, different from the vacuum. It is often assumed that 
open charm photoproduction occurs through the mechanism of the photon-gluon fusion, $\gamma+G\to c+\overline{c}$ \cite{PGF}, even in near threshold region as the applicability of QCD for the considered processes may be justified by the large $c$-quark mass ($m_c\simeq $1.5 GeV). But, if the elementary process 
$\gamma+G\to c+\overline{c}$ can be reliably calculated \cite{PGF}, the application to the exclusive processes $\gamma+N\to Y_c+\overline{D}(\overline{D}^*)$ is not straightforward. 

\begin{figure}
\begin{center}
\includegraphics[width=17cm]{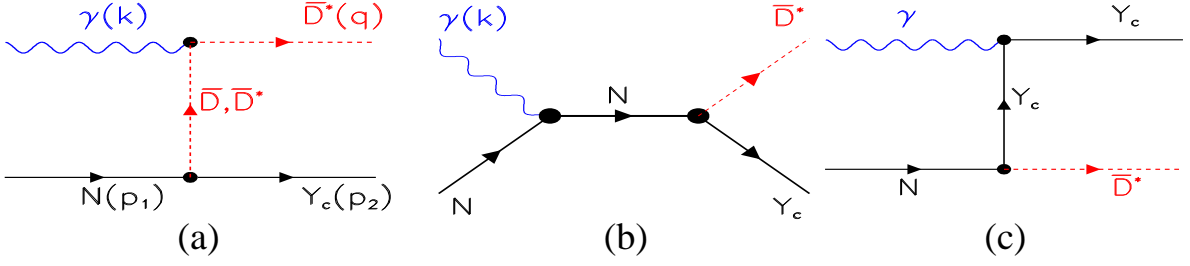}
\caption{\label{fig:fig1} The pole mechanisms for $\gamma+N\to Y_c+\overline{D^*}$.}
\end{center}
\end{figure}

The process $\gamma+N\to Y_c+\overline{D}(\overline{D}^*)$ at large photon energy and at small momentum transfer (i.e. for forward production), can be analzed as the photoproduction of pseudoscalar(vector) light mesons ($\pi$, K) in similar kinematical conditions. In other words the exclusive processes can be described in terms of non-perturbative models, like any binary processes at large values of the total energy $s$ and of small momentum $t$, such as Regge-pole description. So, the large mass of $c$-quarks results in a higher threshold but does not necessarily implies a reaction mechanism of QCD-nature. In this framework, the mechanism based on the elementary subprocess $\gamma+G\to c+\overline{c}$ can be viewed as $D$-meson exchange in $t$-channel, for the process $\gamma+N\to Y_c+\overline{D}^*$ (Fig. \ref{fig:fig1}a). Two other baryon exchanges, the $s$- and $u$-contributions, Figs.\ref{fig:fig1}b and \ref{fig:fig1}c, have to be taken into account to insure the  gauge invariance of the total matrix element: the conservation of electromagnetic current is a very important property of any photoproduction process.

The matrix element corresponding to these diagrams can be written as follows:
\begin{equation}
{\cal M}={\cal M}_t(D)+{\cal M}_t(D^*)+{\cal M}_s+{\cal M}_u.
\label{eq:ms}
\end{equation}
where the indices $t$, $s$ or $u$ indicate the contributions of the corresponding channels.

The exchange by pseudoscalar mesons is described by the following matrix element:
\begin{equation}
{\cal M}_t(D)=ie\displaystyle\frac{\kappa(D^*D\gamma)}{m_v(t-m_D^2)}g_{NY_c\overline{D}}
\overline{u}(p_2)\gamma_5 u(p_1)
\epsilon_{\mu\nu\alpha\beta} \epsilon_{\mu}k_{\nu}U_{\alpha}^*q_{\beta},
\label{eq:td}
\end{equation}
where $\kappa(D^*D\gamma)$ is the transition magnetic moment describing the electromagnetic decay $D^*\to D+\gamma$,  $g_{NY_c\overline{D}}$ is  the  coupling constant for the vertex $N\to Y_c+\overline{D} $, $\epsilon_{\alpha}$$(U_{\alpha})$ is the four vector of the photon $(D^*$-meson) polarization, $m_D$ is the mass of the pseudoscalar $D$-meson, $t=(k-q)^2=(p_1-p_2)^2$. The notation of the particle four momenta is indicated in Fig. \ref{fig:fig1}a.

The constant $\kappa(D^*D\gamma)$ determines the width of the radiative decay 
$D^*\to D\gamma$, through the following formula:
\begin{equation}
\Gamma(D^*\to D\gamma)=\alpha\kappa^2(D^* D\gamma)\displaystyle\frac{m_v}{24}
\left(1-\displaystyle\frac{m_D^2}{m_v^2}\right )^3,
\label{eq:gam}
\end{equation}
where $\alpha=e^2/(4\pi)=1/137$. Evidently the corresponding width does not allow to find the sign of $\kappa(D^*D\gamma)$, which is important for the calculation of possible interference phenomena, for the above quoted observables. However, the quark model gives  indications on this sign, as we will discuss later.

The matrix element ${\cal M}_t(D^*)$, corresponding to vector $D^*$-exchange 
can be written as:
\begin{equation}
{\cal M}_t(D^*)=\displaystyle\frac{e}{t-m_D^2}\overline{u}(p_2)\left [
\gamma_{\alpha}g_1+g_2\displaystyle\frac{\sigma_{\alpha\nu}Q_{\nu}}{m+M}
\right ] u(p_1) 
\left (-g_{\alpha\beta}+ \displaystyle\frac{Q_{\alpha}Q_{\beta}}{m_v^2}\right )
{\cal J}_{\beta},\label{eq:eq18}
\end{equation}
\begin{equation}
{\cal J}_{\beta}=-2\epsilon\cdot q U^*_{\beta}e(\overline{D}^*)+\kappa(\overline{D}^*)\left [ \epsilon_{\beta}k \cdot U^*-\kappa_{\beta}\epsilon\cdot U^*\right ] +(...)
\label{eq:eq19}
\end{equation}
with the following notations:
\begin{itemize}
\item $e(D^*)$ and $\kappa(D^*)$ are the electric charge and the anomalous magnetic moment of the produced $\overline{D}^*$ meson, $e(D^{*0})=0$, $e(D^{*-})=-1$;
\item $Q=p_1-p_2$ is the four momentum of the virtual $D^*$-meson (Fig. \ref{fig:fig1}a);
\item $g_1$  and $g_2$ are the vector and tensor coupling constants for the vertex $N\to Y_c+\overline{D}^*$
\end{itemize}
The term $(...)$ in Eq. (\ref{eq:eq19}) denotes the possible contribution due to the anomalous quadrupole moment of $D^*$, which is generally different from zero, even for the neutral $D^*$-meson - due to the charm content of this meson. For the same reason, $\kappa(D^{*0})\ne 0$. All these electromagnetic constants are not known experimentally, therefore we will neglect in our calculations  possible contributions from the $D^*$-quadrupole moment.

The one nucleon exchange is described by the following matrix element:
\begin{equation}
{\cal M}_s=\displaystyle\frac{e}{s-m^2}\overline{u}(p_2)\left (g_1 \hat U
+g_2\displaystyle\frac{\hat U\hat q}{m+M}\right ) (\hat k+\hat p_1+m)\left [
\hat\epsilon e(N)-\displaystyle\frac{\kappa(N)}{2m}\hat\epsilon \hat k\right ]u(p_1),
\label{eq:eq20}
\end{equation}
where $s=(k+p_1)^2$, $e(N)$ and $\kappa(N)$ are the electric charge and the anomalous magnetic moment of the target nucleon, $e(n)=0$, $e(p)=1$, $\kappa(n)$=-1.91 and $\kappa(p)=1.79 $.

Finally, the matrix element due to the exchange of $Y_c$-hyperon is:
\begin{equation}
{\cal M}_u=\displaystyle\frac{e}{u-M^2}\overline{u}(p_2)\left [
\hat\epsilon e(Y)- \displaystyle\frac{\kappa(Y)}{2M}\hat\epsilon\hat k
\right ]( \hat p_2-\hat k +M) \left ( g_1\hat U+g_2\displaystyle\frac{\hat U\hat q }{m+M}\right )u(p_1),
\label{eq:eq21}
\end{equation}
where $u=(k-p_2)^2$, $e(Y)$ and $\kappa(Y)$ are the electric charge and the anomalous magnetic moment of the
$Y_c$-hyperon: $e(Y)=+1 $ for $\Lambda_c^+$ and $\Sigma_c^+$, $e(Y)=+2 $ for $\Sigma_c^{++}$ and  $e(Y)=0 $ for $\Sigma_c^0$. The anomalous magnetic moment $\kappa(Y)$ is experimentally unknown for any charmed hyperon.

Let us briefly discuss the gauge invariance of the suggested model. All contributions induced by the particle magnetic moments, $\kappa(D^*D\gamma)$, $\kappa(D^*)$, $\kappa(N)$, and $\kappa(Y)$ are automatically gauge invariant, independently on the numerical value of these quantities. For the terms of the matrix element, which are induced by the electrical charges of the particles, the corresponding divergency of the electromagnetic current is determined by the following formula:
\begin{equation}
\Delta {\cal M}={\cal M}(\gamma N\to Y_c\overline{D}^*)\stackrel{\epsilon\to k}{=}
e[e(N)-e(\overline{D}^*)-e(Y)]u(p_2)
\left (g_1 \hat U
+g_2\displaystyle\frac{\hat U\hat q}{m+M}\right )
u(p_1)+ \Delta {\cal M}',
\label{eq:eq22}
\end{equation}
\begin{equation}
\Delta {\cal M}'=\overline{u}(p_2)\left [g_1\displaystyle\frac{m-M}{m_v^2}U\cdot k+g_2
\displaystyle\frac{\sigma_{\alpha\beta}U_{\alpha}k_{\beta}}{m+M}
\right ] u(p_1).
\label{eq:eq23}
\end{equation}
From the conservation of electric charge in the considered reactions: $e(N)=e(\overline{D}^*)+e(Y)$, it follows that
$\Delta {\cal M}=\Delta {\cal M}'\ne 0$, i.e. this part of the matrix element violates the gauge invariance\footnote{The $g_1$-ccontribution to $\Delta {\cal M}'$ can be cancelled by the "catastrophic" diagram. The corresponding predictions \protect\cite{Ru78} for the cross sections of the processes $\gamma +N\to Y_c+\overline{D}(\overline{D}^*)$, in the threshold region, are in contradiction with the experimental data \protect\cite{Ab84}.}.

In conditions of collinear kinematics, however, the situation with gauge invariance is essentially simplified. To show this, let us consider an 'improved' matrix element, made gauge invariant by the following substitution:
\begin{equation}
{\cal M}\to {\cal M}'={\cal M}-\displaystyle\frac{\epsilon\cdot\overline{p}}{k\cdot\overline{p}}
\Delta {\cal M}',
\label{eq:eq24}
\end{equation}
where $\overline{p}$ is a four vector, built generally as a linear combination of the particle four-momenta:
\begin{equation}
\overline{p}=ak+bp_1+cq,
\label{eq:eq25}
\end{equation}
where the coefficients $a$, $b$, and $c$ are arbitrary functions of the Mandelstam variables $s$ and $t$. With this transformation the resulting matrix element, ${\cal M}'$ is gauge invariant. But in collinear kinematics $\epsilon\cdot\overline{p}$=0, for any choice of $a$, $b$, and $c$, so, when  $\vec \epsilon$ is transverse,
$\vec\epsilon\cdot\vec k$=0, we find  ${\cal M}' $=${\cal M} $. Therefore, all calculations using this transverse gauge condition, which holds in the reaction CMS, can be done on the basis of Eqs. (\ref{eq:td}),
(\ref{eq:eq19}), (\ref{eq:eq20}), and (\ref{eq:eq21}), for different matrix elements, without violating gauge invariance. 

Note in this respect, that the contribution to the amplitudes $f_i$ which is proportional to the $D^*$ electric charge, $e(D^*)$ vanishes in collinear kinematics, see Eq. (\ref{eq:eq19}).

The explicit expressions for the collinear amplitudes $f_i$, $i=1,2,3$,  corresponding to the different matrix elements, see Eqs. (\ref{eq:td}), (\ref{eq:eq19}), (\ref{eq:eq20}), and (\ref{eq:eq21}), are given in the Appendix. In the considered model, these amplitudes are real functions of $E_{\gamma}$.

\section{The electromagnetic coupling constants and the form factors.}
We consider here the six possible binary reactions of $D^*$-photoproduction:
\begin{eqnarray}
&\gamma+p \to \Lambda_c^+ +\overline{D}^{*0}, &\gamma+n \to \Lambda_c^+ +D^{*-},\nonumber\\
&\gamma+p\to \Sigma_c^+  + \overline{D}^{*0},&\gamma+n \to \Sigma_c^+  + D^{*-}, \label{eq:eq27}\\
&\gamma+p\to\Sigma_c^{++}+D^{*-}, &\gamma+n \to \Sigma_c^0  + \overline{D^{*0}}\nonumber
\end{eqnarray} 
and calculate the $E_{\gamma}$-dependence of the following observables: $d\sigma/dt $, $\rho_{11}$, $A_z$ and $P_z$ for each of the reactions (\ref{eq:eq27}).

The corresponding collinear amplitudes are linear functions of the strong coupling constants for the two vertices, $N\to Y_c+\overline{D}$ and $N\to Y_c+\overline{D}^*$. So, taking into account the isotopic invariance of the strong interaction, it is necessary to know at least six independent coupling constants:
\begin{eqnarray}
&g_{1\Sigma}\equiv g_1(p\Sigma_c^+\overline{D}^{*0}), 
&g_{1\Lambda}\equiv g_1(p\Lambda_c^+\overline{D}^{*0}),\nonumber\\
&r_{12}(\Sigma)\equiv g_2(p\Sigma_c^+\overline{D}^{*0})/g_1(p\Sigma_c^+\overline{D}^{*0}), 
&r_{12}(\Lambda)\equiv g_2(p\Lambda_c^+\overline{D}^{*0})/g_1(p\Lambda_c^+\overline{D}^{*0}), \label{eq:eq28}\\
&r(\Sigma)= g(p\Sigma_c^+\overline{D}^{0})/g_1(p\Sigma_c^+\overline{D}^{*0}), 
&r(\Lambda)\equiv g(p\Lambda_c^+\overline{D}^{0})/g_1(p\Lambda_c^+\overline{D}^{*0}),\nonumber
\end{eqnarray} 
So, the two possible processes of $\Lambda_c^+$-photoproduction, $\gamma+N\to \Lambda_c^++\overline{D}^{*}$, can be described by a set of three coupling constants:
$$g_{1\Lambda},~r_{12}(\Lambda),\mbox{~and}~r(\Lambda)$$
and the four possible reactions of $\Sigma_c$ production, $\gamma+N\to \Sigma_c+\overline{D}^{*}$, can be described by another set of coupling constants:
$$g_{1\Sigma},~r_{12}(\Sigma),\mbox{~and}~r(\Sigma).$$

All polarization observables depend on two ratios only, $r_{12}(Y)$ and $r(Y)$ , and are independent on the electric charges of the initial nucleon and the produced charmed particles, due to isotopic invariance. The coupling constant $g_{1\Lambda}$($g_{1\Sigma}$) is important for the prediction of the absolute value of the differential cross section, with the following isotopic relations:
$$
g_1^2(p\Lambda_c^+\overline{D}^{*0})=g_1^2(n  \Lambda_c^+ D^{*-}),$$
\begin{equation}
g_1^2(p\Sigma_c^{++}D^{*-})=g_1^2(n \Sigma_c^0 \overline{D}^{*0})=2g_1^2(p\Sigma_c^+\overline{D}^{*0})
=2g_1^2(n\Sigma_c^+\overline{D}^{*-}).
\label{eq:eq29}
\end{equation}

But the electromagnetic properties of the nucleon and the charm particles, which strongly depend on the 
electric charge of the particles, induce large isotopic effects, i.e. a strong dependence on the type of reaction. So, for the numerical estimations it is necessary to know the following magnetic moments of charmed baryons and $D^*\to D+\gamma$ electromagnetic transitions:
$\kappa(Y_c)$, $\kappa(D^{*-}D^{*-}\gamma), $~ $\kappa(D^{*0}D^0\gamma)$.

The quark model gives  prescriptions which relate the magnetic moments of the charmed hyperons to the magnetic moments of the charmed quarks, $\mu_q$, $q=u$, $d$, or $c$:
\begin{eqnarray}
\mu(\Lambda_c^+)&= &\mu_c,\nonumber\\
\mu(\Sigma_c^{++})&=&(4 \mu_u-\mu_c)/3,\label{eq:eq30}\\
\mu(\Sigma_c^+)&=&(2 \mu_u+2\mu_d-\mu_c)/3,\nonumber \\
\mu(\Sigma_c^0)&=&(4\mu_d-\mu_c)/3.\nonumber
\end{eqnarray} 
The magnetic moments of point-like quarks are determined by the electric charge 
of the quark and its mass, so
\begin{equation}
\mu_q=\displaystyle\frac{Q_q}{2m_q},
\label{eq:eq31}
\end{equation}
where $m_q$ is the 'constituent quark' mass. From the analysis of the nucleon and the usual hyperons magnetic moments, one finds \cite{PDG}:
\begin{equation}
\mu_u=1.852~\mu_N,~\mu_d=-0.9722~\mu_N,
\label{eq:eq32}
\end{equation}
where $\mu_N$ is the nucleon magneton. Using these values, one has $m_u=338$ MeV, and $m_d=322$ MeV. Therefore, the current value of $m_c=1.5$ GeV for the charm quark results in  $\mu_c=0.42$ $\mu_N$, and in the values of the  magnetic moments and anomalous magnetic moments reported in Table \ref{tab:tab1}.
\begin{table}[h]
\begin{tabular}{|c|c|c|}
\hline\hline
Particle &$\mu(Y) $ &$\kappa(Y)$\\
\hline\hline
$\Lambda_c^+$& 0.42 & -0.58 \\
\hline
$\Sigma_c^{+}$&0.45&-0.55 \\
\hline
$\Sigma_c^{++}$&2.33&0.33 \\
\hline
$\Sigma_c^0$&-1.44&-1.44 \\
\hline\hline
\end{tabular}
\caption{ Magnetic moments and anomalous magnetic moments ($\kappa(Y)=\mu(Y)-e(Y)$) of  charmed baryons, in units of $\mu_N$, in framework of the quark model.}
\label{tab:tab1}
\end{table}

For completeness, let us mention that other prescriptions exist for the calculation of $\kappa(Y)$: SU(4)-symmetry \cite{Re1}, bag models \cite{Bo80}, different versions of quark model \cite{Je86}, dispersion sum rules \cite{Sa94}, etc.

In principle, the reactions $\gamma+ N\to Y_c +\overline{D}^*$ can be considered a possible source of information on charm hyperon magnetic moments. In the literature, another possibility of measuring the magnetic moment of  $\Lambda_c^+$, based on the precession in bending crystals has been discussed \cite{Sa96}, but this method can not be applied to the $\Sigma_c$-hyperons, which main decay, $\Sigma_c\to \Lambda_c+\pi$ occurs through the strong interaction.

The quark model can also be used for the prediction of the transition magnetic moments (again in terms of quark magnetic moments):
$$\displaystyle\frac{\kappa(D^{*-}D^{-}\gamma)}{\kappa(D^{*0}\overline{D}\gamma)}=
\displaystyle\frac{\mu_c+\mu_d}{\mu_c+\mu_u}\simeq -0.24.$$
Taking the existing experimental data about  $D^{*0}$, $Br(D^{+0}\to D^{+}\gamma)=(1.6\pm 0.4)$\%, and $\Gamma_T(D^{*+})=(96\pm 22)$ keV \cite{PDG}, one can find from Eq. (\ref{eq:gam}): $|\kappa(D^{*-}\overline{D}\gamma)|\simeq 1$. Moreover, the quark model allows to fix the sign of this magnetic moment, as $\kappa(D^{*-}\overline{D}\gamma)>0$. 

This allow us to fix all the necessary electromagnetic constants, i.e. the absolute values of the magnetic moments and their signs. The signs, in particular, are very important in the analysis of the isotopic effects for these reactions, which are large, in the present model, near threshold, due to the strong interference of the different contributions.

Finally, to fix the strong coupling constants, we use, as in Ref. \cite{Etg03}, the existing information on the corresponding coupling constants for strange particles, which can be determined from the analysis of the data on photo and electroproduction of $\Lambda$ and $\Sigma$-hyperons on protons. We will take the following values \cite{Gu97}:
\begin{eqnarray}
g_1(N\Lambda K^*)=&-23.0,&r_{12}(\Lambda)=2.5,\nonumber \\
g_1(N\Sigma^0 K^*)=&-25.0,&r_{12}(\Sigma)=-1.0,\label{eq:eq33}\\
g^2_{K\Lambda N}/4\pi=&10.6,&g^2_{K\Sigma N}/4\pi=1.6,\nonumber
\end{eqnarray}
with $g_{K\Lambda N}<0$ and $g_{K\Sigma N}>0$ in agreement with SU(3) constrains. The same holds also for $K^*$-coupling constants. Using SU(4) symmetry, i.e. the substitution $s\to c$, one can find from (\ref{eq:eq33}) the necessary coupling constants, for charm particles.

The last problem of the model to be discussed is the parametrization of the phenomenological form factors, which are important ingredients of ELA approaches and are usually introduced for the pole contributions \cite{Re03a,Ga02}. At each vertex, for $D$ and $D^*$-exchanges, one can parametrize the form factor as: 
\begin{equation}
F_{1,2}(t)=\displaystyle\frac{\Lambda_{1,2}^2-m_D^2}{\Lambda_{1,2}^2-t},~F_{1,2}(t=m_D^2)=1,
\label{eq:eq34}
\end{equation}
where the index 1(2) corresponds to the electromagnetic (strong) vertex and $\Lambda_{1,2}$ is the corresponding cut-off parameter. The baryon contributions, Eqs. (\ref{eq:eq20}) and (\ref{eq:eq21}), have to be modified by the following form factors:
\begin{eqnarray}
&F_N(s)=\displaystyle\frac{\Lambda_N^4}{\Lambda_N^4+(s-m^2)^2},&\mbox{~for ~s-channel},\nonumber \\
&F_Y(u)=\displaystyle\frac{\Lambda_Y^4}{\Lambda_Y^4+(u-m^2)^2},&\mbox{~for ~u-channel},\label{eq:eq35}
\end{eqnarray}
where  $\Lambda_N $ and $\Lambda_Y $ are the corresponding cut-off parameters, generally different from $\Lambda_{1,2}$.

Let us note that the introduction of  these form factors violates the gauge invariance of that part of the matrix element, ${\cal M}$, which is determined by the electric charges of the participating hadrons. Again let us mention here, that in case of collinear kinematics, the nonconserving contributions cancel for the transverse gauge of the photon polarization, as one can see from the substitution (\ref{eq:eq24}).

It follows, that in collinear regime, we can take the parametrizations (\ref{eq:eq34}) and (\ref{eq:eq35}), with a different form factor for each diagram and with independent values of the cut-off parameters, without violation of the gauge invariance.

\section{Discussion of numerical predictions}

Using the expressions for the collinear amplitudes $f_i$, $i=1-3$, (see Appendix) and the numerical values for the strong and electromagnetic couplings given above, we can predict the $E_{\gamma}$-behavior of the differential cross section, $d\sigma/dt$ and of some polarization observables for the collinear regime. We will consider, more exactly, only forward production. In this kinematics, the contributions of baryonic exchanges are negligible, rapidly decreasing with energy, in comparison with $t$-channel contributions - due, on one side, to the effect of the two form factors, $F_N(s)$ and $F_Y(u)$, and, on another side, to the relative size of the $t$, $s$, and $u$ propagators. Therefore in the numerical estimations, the $t$-channel contribution dominates. Assuming, for simplicity, a common form factor for the $D+D^*$-contributions:
$$F(t)=\left (\displaystyle\frac{\Lambda^2-m_D^2}{\Lambda^2-t}\right )^2,$$
corresponding to $\Lambda_1=\Lambda_2=\Lambda$, we take $\Lambda$ as a free parameter, to be adjusted on the data.

\begin{figure}
\begin{center}
\includegraphics[width=17cm]{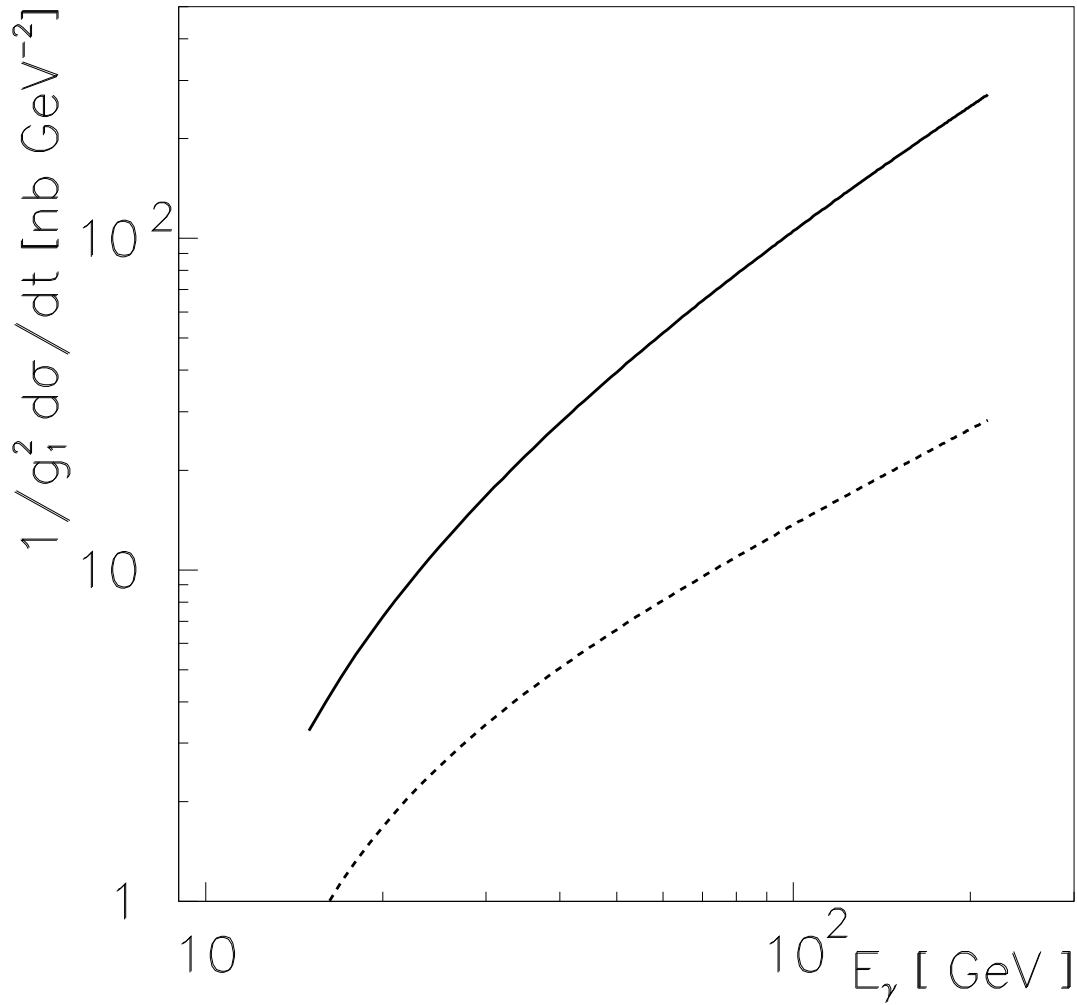}
\caption{\label{fig:fig2} Reduced differential cross section,  as a function of $E_{\gamma}$, for  $\gamma+p\to \Lambda_c^+ 
+\overline{D^{*0}}$ (solid line), and 
$\gamma+p\to\Sigma_c^{++}+D^{*-}$ (dashed line).}
\end{center}
\end{figure}

For an exponential $t$-dependence of the cross section, for $\gamma+N\to Y_c+\overline{D}^*$-processes:
$$\displaystyle\frac{d\sigma}{dt}=\left( \displaystyle\frac{d\sigma}{dt}\right )_{\theta=0}e^{b(t-t_{max})},
$$
with $t_{max}=m_v^2-2k(E_v-q)$ ($q$ is the three-momentum of the produced $\overline{D}^*$), one can find for the total cross section:
$$\sigma(\gamma N\to Y_cD^*)=\displaystyle\frac{1}{b}\left( \displaystyle\frac{d\sigma}{dt}\right )_{\theta=0}.
$$
Without direct experimental information about the cross section for the processes $\gamma +N\to Y_c+ \overline{D}^*$, we take a conservative assumption:
$$\sigma(\gamma N\to Y_c\overline{D}^*)\simeq 0.1\sigma_T(\gamma N\to \mbox{open~charm}),$$
in the interval $E_{\gamma}=100\div 200$ GeV. For $b\simeq 5$ GeV$^{-2}$, according to the analysis \cite{Sl83},
we find  that
\begin{equation}
\left( \displaystyle\frac{d\sigma}{dt}\right )_{\theta=0}
\simeq  b\sigma_T(\gamma N\to  \mbox{open~charm})\simeq ~250 \mu \mbox{b~ GeV}^{-2},
\label{eq:eq36}
\end{equation}
at $E_{\gamma}=200$ GeV. Then, we fix $\Lambda \simeq$ 2.5 GeV, in agreement with the value previously used for the calculation of associative charm production, $N+N\to N+Y_c+\overline{D}$, near threshold \cite{Re03a,Ga02}.
The differential cross section, $d\sigma/dt$ (more exactly, the reduced differential cross section, $1/g_1^2~ d\sigma/dt$) is  shown  as a function of $E_{\gamma}$, in Fig. \ref{fig:fig2}, for two reactions, $\gamma+p \to \Lambda_c^+ +\overline{D}^{*0}$ (solid line) and $\gamma+p\to\Sigma_c^{++}+D^{*-}$ (dashed line). Note that the reduced cross section does not depend on the constant $g_1$, which is different for the different reactions. Therefore we report the results for two reactions, only. 

\begin{figure}
\begin{center}
\includegraphics[width=17cm]{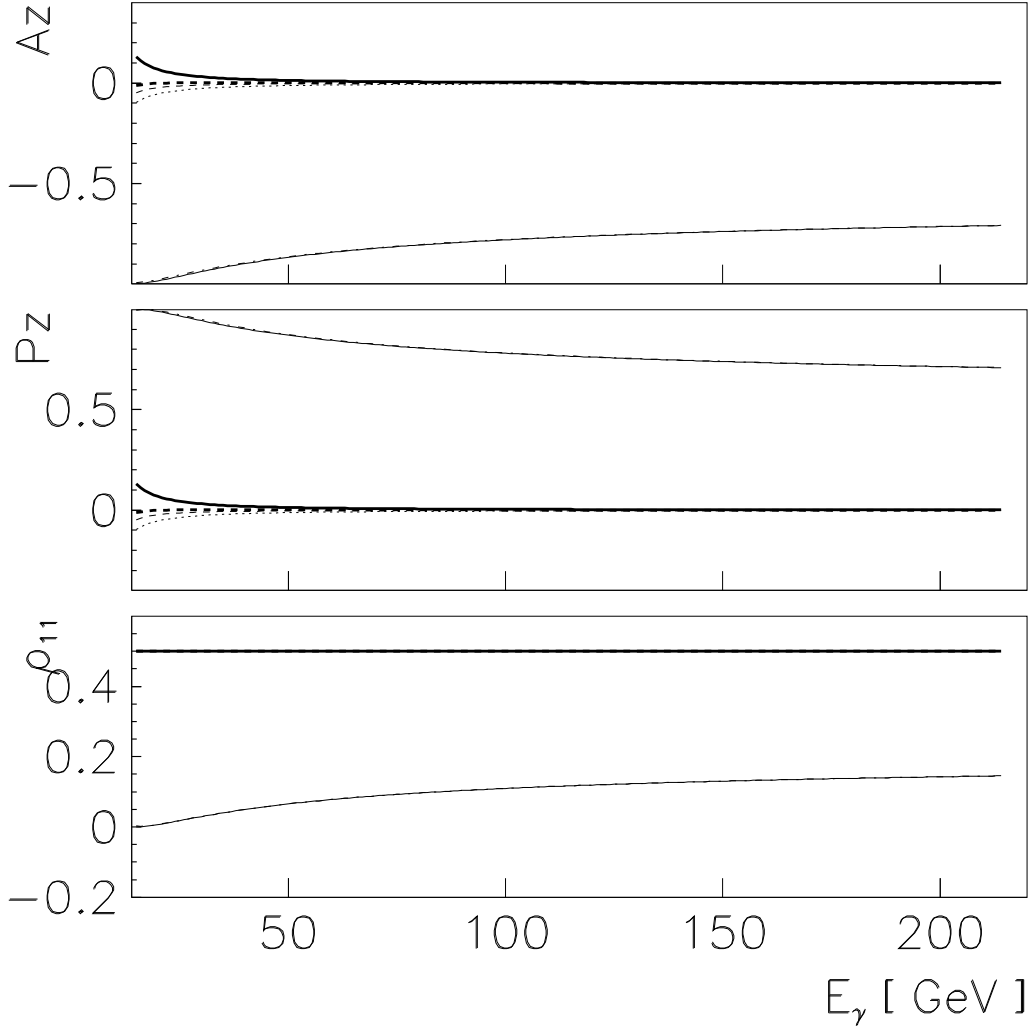}
\caption{\label{fig:fig3} Different polarization observables: asymmetry $A_z$ (top), polarization $P_z$ (center) and $\rho_{11}$ (bottom), for the six considered reactions as a function of $E_{\gamma}$:
$\gamma+p \to \Lambda_c^+ +\overline{D}^{*0}$ (solid line),  
$\gamma+p\to \Sigma_c^+  + \overline{D}^{*0}$ (dashed line),
$\gamma+p\to\Sigma_c^{++}+D^{*-}$ (dotted line),
$\gamma+n \to \Lambda_c^+ +D^{*-}$ (dash-dotted line), 
$\gamma+n \to \Sigma_c^+  + D^{*-}$ (thick solid line), 
$\gamma+n \to \Sigma_c^0  + \overline{D^{*0}}$ (thick dashed.}
\end{center}
\end{figure}

One can see that, in the considered model, the  $\Lambda_c$-production has the largest reduced  cross section.  The fact that all four reactions $\gamma+N\to\Sigma_c+\overline{D}$ are one order of magnitude smaller, is due to the difference between the values of $r_{12}(Y)$ for the $N\to \Lambda_c\overline{D}^*$ and $N\to \Sigma_c\overline{D}^*$-vertices: the 'magnetic' combination $g_1+ g_2$ of the corresponding coupling constants cancels for $r_{12}(\Sigma)=-1$.  This is also the reason of the difference of the polarization observables for $\Lambda_c$ and $\Sigma_c$-photoproduction (Fig. \ref{fig:fig3}). Note that all polarization observables do not depend on the form factor $F(t)$ and on the coupling constant $g_1$.

However, the absolute values of the cross section for any process $\gamma+N\to Y_c+\overline{D}^*$, calculated with the coupling constants $g_1$, Eq. (\ref{eq:eq33}), are too large, in comparison with the expectation (\ref{eq:eq36}). This means that SU(4) symmetry is strongly violated, with respect to these constants, in agreement with a previous observation \cite{Re01}.

The density matrix element is almost independent on energy (Fig. \ref{fig:fig3}c) , and $\rho_{11}\simeq 0.5$ for $\gamma +N\to \Lambda _c+ \overline{D}^*$, whereas $\rho_{11}\le 0.2$ for $\gamma +N\to \Sigma_c+\overline{D}^*$ . The maximal value of 
$\rho_{11}$ produces a $\sin^2\theta$-distribution of the produced $D$-meson, through the decay $D^*\to D+\pi$ ($\theta $ is the angle between the $\vec k$-direction and and the direction of the $D$-meson three momentum in the $D^*$-rest system). Such $\theta$-dependence results in a depletion of $D$-meson production at small angles, which should be observed, for example, in the COMPASS experiment.

The large (in magnitude) and negative values of the asymmetry $A_z$, for $\vec\gamma+\vec N\to \Lambda_c^++\overline{D}^*$, are near the limiting value $A_z=-1$; more exactly, $|A_z|\ge$ 0.8, for $E_{\gamma}\le$ 100 GeV. 

If one considers this reaction as a background for the PGF mechanism, one can estimate the effect of this result on the extraction of $\Delta G$. The PGF asymmetry, $A_{PGF}$, is predicted by perturbative QCD, on the basis of the hard subprocess $\gamma+G\to c+\overline{c}$ \cite{PGF}, and can be related to the gluon contribution to the nucleon spin. The contribution of the exclusive process $\vec\gamma+\vec N\to Y_c+\overline{D}^*$ to the asymmetry, can be parametrized as follows:
\begin{equation}
A(\vec\gamma \vec N\to \mbox{charm}+X)=\displaystyle\frac{A_{PGF}+RA_z(\gamma  N\to Y_c \overline{D}^*)}{1+R},
\label{eq:eq36a}
\end{equation}
where $R=\sigma(\gamma  N\to Y_c \overline{D}^*)/\sigma(\gamma  N\to\mbox{charm}+X)$, neglecting, for simplicity, other sources of background due to additional channels of charm particle photoproduction. In case of $R\ll 1$, we can write:
\begin{equation}
A(\vec\gamma \vec N\to \mbox{charm}+X)=A_{PGF}+\Delta A, ~\Delta A=R[A_z(\gamma  N\to Y_c \overline{D}^*)-A_{PGF}]
\label{eq:eq37}
\end{equation}
One can see, that in case of opposite signs of the two asymmetries, the 'dilution factor', $1/(1+R)$ and the background $A_z$-contribution, due to the channel $\gamma  +N\to Y_c +\overline{D}^*$, act coherently in increasing the correction $\Delta A$ (in absolute value). For example, if $R\simeq 0.1$, $A_{PGF}\simeq 0.3 $ and $A_z(\gamma  N\to Y_c \overline{D}^*)\simeq -1$, one can find $\Delta A\simeq -0.13$, which represents a correction $\delta A/A_{PGF}\simeq 43$\%. This is a large correction, even for a relatively small contribution of the considered exclusive channel to the total cross section.

\begin{figure}
\begin{center}
\includegraphics[width=17cm]{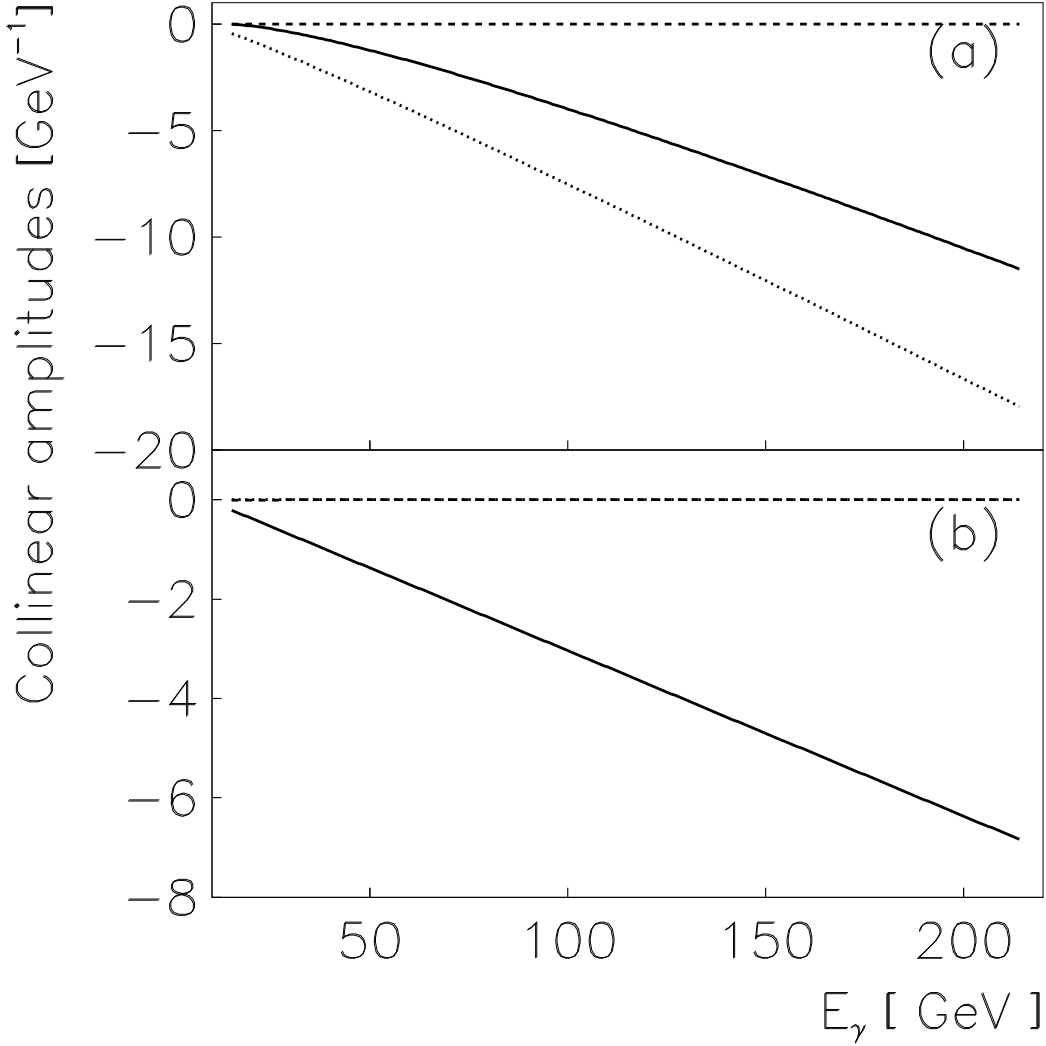}
\caption{\label{fig:fig4} Collinear amplitudes $f_1$ (solid line), $f_2$ (dashed line), and $f_3E_v/m_v$ (dotted line), as functions of $E_{\gamma}$ for  $\gamma+p\to \Lambda_c^+ +\overline{D^{*0}}$ (a) and 
$\gamma+p\to\Sigma_c^{++}+D^{*-}$ (b).}
\end{center}
\end{figure}

The energy behavior of the collinear amplitudes, $f_1$, $f_2$ and $f_3E_v/m_v$, for the different channels 
 $\gamma  N\to Y_c \overline{D}^*$, is shown in Fig. \ref{fig:fig4}, taking into account the form factors for the $s+u+t$-contributions. Due to the fact that the baryonic contributions are small, the relative values of these amplitudes are independent on the form factor. The specific factor, $E_v/m_v$, which has been introduced for the amplitude $f_3$, results from the relativistic description of the $D^*$-meson polarization properties. The collinear amplitude $f_3$ describes the $D^*$-production with longitudinal polarization, $\vec U\cdot\hat{\vec k}\ne 0$, and the $z$-component of such polarization has to be equal to $E_v/m_v$, due to the orthogonality condition, $U\cdot q$=0.

Note that the different contributions, taken into account in the present model, have very simple and transparent polarization properties, especially in collinear kinematics. For example, the pseudoscalar $\overline{D}$-exchange, which induces a single collinear amplitude, $f_2$, results in the following values of the polarization observables:
\begin{equation}
\rho_{11}=1/2,~A_z=P_z=0,~D_{zz}=1,~D_{xx}=D_{yy}=0.
\label{eq:eq38}
\end{equation}
These numbers reflect the fact that a zero-spin exchange does not transfer any information on the polarization from the electromagnetic to the strong vertex. 
The $D^*$-vector exchange, which is characterized by a nonzero anomalous magnetic moment $\kappa(D^*)$, results in $f_2=0$. This means that the $D\bigotimes D^*$-interference vanishes for $d\sigma/dt$ and $\rho_{11}$. Note that $A_z=-P_z$ for $D^*$-exchange only, independently on the photon energy $E_{\gamma}$. Therefore, the inequality $A_z+P_z\ne 0$ characterizes the size of the $D\bigotimes D^*$-interference, being sensitive to the sign of the ratio $r(Y)$.

The $N-$exchange in $s-$channel gives $f_1=f_2$, which results in $A_z=-1$,  independently on the reaction channel, photon energy and numerical values of the coupling constants. This result follows from the helicity properties of $s$-channel, where the spin 1/2 in the intermediate state forbids the helicity transition $\pm 3/2 \to \pm 3/2 $, which is characterized by the combination of collinear amplitudes $f_1-f_2$. But $\rho_{11}$ and $P_z$ can take any value in the allowed limits:
$$0\le\rho_{11}\le1/2,~0\le |P_z|\le 1.$$

Finally, for the $u$-channel contribution (with $f_1=-f_2$), the discussed polarization observables can take any value, but $P_z=-1$, for all reactions and at any photon energy.

Polarization phenomena depend strongly on the relative role of these contributions and may show large sensitivity to the value of the strong and electromagnetic coupling constants.

\section{Conclusions}

We considered the exclusive photoproduction of charmed vector mesons, $\gamma+N\to Y_c+\overline{D}^*$, in collinear kinematics, where the differential cross section is large and the spin structure of the matrix element is essentially simplified. We analyzed firstly the single and double-spin polarization phenomena, in a general form, in terms of three independent collinear amplitudes. The energy dependence of these amplitudes has been predicted - for the six possible processes, in a QCD-inspired model, which is the basis of an effective Lagrangian approach. The strong coupling constants for the vertices $N\to Y_c+\overline{D}$ and $N\to Y_c+
\overline{D}^*$ can be related through SU(4) symmetry with the corresponding coupling constants for strange particles, i.e. for the vertices $N\to Y+K$ and $N\to Y+K^*$, $Y=\Lambda$ or $\Sigma$-hyperon, which are known from the analysis of experimental data concerning photo-and electroproduction of strange particles. 

The electromagnetic characteristics of charmed particles, such as the magnetic moments of the charmed $Y_c$-hyperons and the transition magnetic moments for the decays $D^*\to D+\gamma$ have been estimated in framework of the quark model.

As the baryonic exchanges (by $N$ in $s$-channel and $Y_c$ in $u$-channel) are negligible in the considered model, the polarization observables are quite insensitive to this form factor, and to the vector coupling constant $g_1(NY_c\overline{D}^*)$ as well. Therefore this model gives robust predictions for the polarization effects. The large and negative values of the asymmetry $A_z$ (for the collision of a circularly polarized photon beam with a longitudinally polarized nucleon target)  for $\Lambda_c^+$-production on proton and neutron targets, (which is a factor ten larger in comparison of $\Sigma_c$-production), can be source of large systematic error in the extraction of $\Delta G$ from $\vec\gamma+\vec N\to$ charm$+X$, even in case of a relatively small cross section of the considered exclusive reactions.

A reasonable value for the cross section of the considered processes can be obtained only for a smaller value of the coupling constants $g_1(NY_cD^*)$, in comparison with SU(4) predictions.

\section{Acknowledgments}

We thank the members of the Saclay group of the COMPASS collaboration,for interesting discussions and useful comments.

\section{Appendix}

Here we give the expressions for the scalar amplitudes $f_i$, $i=1-3$:
$$ f_i=f_{i,t}(D)+f_{i,t}(D^*)+f_{i,s}+f_{i,u},$$
where the indices $s$, $u$, and $t$ correspond to $s$, $u$, and $t$ channel contributions.
\begin{itemize}

\item $t$-channel ($D$-contribution):
\begin{eqnarray*}
f_{1,t}(D)&=&f_{3,t}(D)=0,\\
f_{2,t}(D)&=&\kappa(D^{*0}\overline{D}\gamma)\displaystyle\frac{g(Y_c)}{2m_v}
\displaystyle\frac{t-m^2_v}{t-m^2_D}\left (1-Q-\displaystyle\frac{2m}{W+m}\right )
\end{eqnarray*}

\item $t$-channel ($D^*$-contribution):
\begin{eqnarray*}
f_{1,t}(D^*)&=&\displaystyle\frac{\kappa(D^*)}{2}
\left (\displaystyle\frac{m-M}{m^2_v}+ \displaystyle\frac{r_{12}}{m+M}
 \right ) +\kappa(D^*)W R_D\left [ (1+r_{12})(1+Q)-r_{12}\displaystyle\frac{W+m}{M+m}\right ],\\
f_{2,t}(D^*)&=&0,\\
f_{3,t}(D^*)&=&-\displaystyle\frac{\kappa(D^*)}{2}(1+r_{12})R_D
\left [W-m+(W+m)Q\right ],
\end{eqnarray*}
with ${R_D}=\displaystyle\frac{E_v+q}{m^2_v(W-m)}.$
\item $s$-channel:

\begin{eqnarray*}
&f_{1,s}=f_{2,s}=&\displaystyle\frac{e(N)}{W+m}(1+Q)+\displaystyle\frac{\kappa(N)}{2m}
\left (Q-1+\displaystyle\frac{2m}{W-m}\right )\\
&&-\displaystyle\frac{e(N)}{W+m}\displaystyle\frac{r_{12}}{m+M}
\left [(W-m)-Q(W+m) \right] \\
&&+\displaystyle\frac{\kappa(N)}{2m}\displaystyle\frac{r_{12}}{m+M}
\left [(W-m)\left(1+\displaystyle\frac{2m}{W+m}\right )+Q(W+m) \right] 
\end{eqnarray*}

\begin{eqnarray*}
&f_{3,s}=&\displaystyle\frac{e(N)}{W+m}(-1+Q)\left( 1-\displaystyle\frac{q}{q_0}\right )\\
&&+\displaystyle\frac{\kappa(N)}{2m}
\left \{ Q+1-\displaystyle\frac{2m}{W-m}+\left [1+\left(1- \displaystyle\frac{2m}{W+m}\right )Q\right ]
\displaystyle\frac{q}{q_0}\right \}\\
&&+\displaystyle\frac{e(N)}{W+m}\displaystyle\frac{r_{12}}{m+M}
\left( 1-\displaystyle\frac{q}{q_0}\right )\left [(W-m)+Q(W+m) \right] \\
&&+\displaystyle\frac{\kappa(N)}{2m}\displaystyle\frac{r_{12}}{m+M}
-\left \{(W+m) Q -(W-M)\left(1- \displaystyle\frac{2m}{W+m}\right )\right .\\
&&\left.  \left [W-m-Q(W+m)\left(1-
 \displaystyle\frac{2m}{W+m}\right )\right ]\displaystyle\frac{q}{q_0}\right\}.
\end{eqnarray*}
\item  $u$-channel :
\end{itemize}
\begin{eqnarray*}
f_{1,u}&=&f_{2,u}=R_u\left \{ e(Y)+\displaystyle\frac{\kappa(Y)}{2M}
\left [M-(q+E_2)\displaystyle\frac{m}{W}\right ]  +e(Y)\displaystyle\frac{r_{12}}{m+M}(q_0-q) \displaystyle\frac{m}{W}+\right .
\\
&& \left . \displaystyle\frac{\kappa(Y)r_{12}}{M(m+M)}(q_0+q)\left [-q-E_2+m\displaystyle\frac{m}{W}
\left (\displaystyle\frac{q_0-q}{q_0+q}\right) \right] 
\right \}\\
f_{3,u}&=&R_u\left \{e(Y)\displaystyle\frac{m}{W}\left( 1-\displaystyle\frac{q}{q_0}\right )\displaystyle\frac{\kappa(Y)}{2m}
\left [-(q+E_2)\left( 1+\displaystyle\frac{q}{q_0}\right )
+m\displaystyle\frac{M}{W}\left( 1-\displaystyle\frac{q}{q_0}\right )\right] \right .\\
&&+e(Y)\displaystyle\frac{r_{12}}{m+M}\displaystyle\frac{m_v^2}{q_0}\\
&&\left .+\displaystyle\frac{\kappa(Yr_{12})}{M(m+M)}(q+q_0)\left [ M-(q+E_2)\displaystyle\frac{m}{W}
 \left (\displaystyle\frac{q_0-q}{q_0+q} \right )- \left (M+(q+E_2)\displaystyle\frac{m}{W}\left (
\displaystyle\frac{q_0-q}{q_0+q} \right )\right )
\displaystyle\frac{q}{q_0}\right]\right \}
\end{eqnarray*}

with $R_u=(1+Q)\displaystyle\frac{W(E_2-q)}{M^2(W+m)}$,  $Q=\displaystyle\frac{q}{E_2+M}$,
$~E_2=\displaystyle\frac{s+M^2-m_v^2}{2w}$, and $q_0=W-E_2$.

{}

\end{document}